\documentclass[aps,prl,twocolumn,showpacs]{revtex4}
\usepackage{graphicx}
\begin{document}

\title{No-core shell model in an effective-field-theory framework}

\author{I. Stetcu}

\author{B.R. Barrett}

\author{U. van Kolck}

\affiliation{Department of Physics, University of Arizona, 
Tucson, Arizona 85721}

\date{September 11, 2006}

\begin{abstract}
We present a new approach to the construction
of effective interactions suitable for many-body calculations 
by means of the no-core shell model (NCSM).
We consider an effective field theory (EFT) with only nucleon fields
directly in the NCSM model spaces.
In leading order, we obtain the strengths of the three 
contact terms from the condition that in each model space 
the experimental ground-state energies of $^2$H, $^3$H and $^4$He
be exactly reproduced. 
The first $(0^+;0)$ excited state of $^4$He and the ground state of $^6$Li 
are then obtained by means of NCSM calculations
in several spaces and frequencies. 
After we remove the harmonic-oscillator frequency 
dependence, we predict for $^4$He an energy level for the first $(0^+;0)$ 
excited state in remarkable agreement 
with the experimental value. The corresponding $^6$Li binding energy 
is about 70\% of the experimental value,
consistent with the expansion parameter of the EFT.
\end{abstract}
\maketitle

The microscopic description of quantum systems, from nuclei to atoms and 
molecules, is a very difficult task that often involves solving numerically 
the many-body Schr\"odinger equation in a restricted space. 
A longstanding challenge has been the construction of 
effective interactions within the finite model space 
where
the many-body problem is solved \cite{effOp98}.
In order to avoid uncontrolled approximations often invoked,
we present a novel approach based on the 
general principles of effective field theory (EFT), which relates intimately 
the effective interactions (and, in general, effective operators
describing
interactions with external probes) 
with the 
model space.

The no-core shell model (NCSM) is a powerful many-body technique 
that provides the solution to the Schr\"odinger equation for 
$A$ interacting nucleons in a restricted space 
\cite{Navratil:2000ww}. 
The NCSM basis states 
are constructed using harmonic-oscillator (HO) wave functions.
Starting from 
two-nucleon (NN) interactions that accurately fit the experimental phase 
shifts and deuteron properties, 
and theoretical three-nucleon (3N) forces adjusted to reproduce 
triton properties, one generally uses a unitary transformation
approach to obtain effective interactions in a restricted model space, 
where an exact diagonalization in a many-body basis can be performed. 
This approach has been very successfully applied to the description of 
energy spectra of  $^4$He and light nuclei throughout the $p$ shell
\cite{Navratil:2000ww,Navratil:2001},
and even beyond \cite{Vary:2006nw}. 
However, the results remain dependent on the details of the 
two- and three-body interactions  used as input. 
Moreover, the method involves the 
``cluster approximation,'' 
which neglects higher-body correlations in a less controlled way. 
Finally, 
at the two-body-cluster level low-momentum 
observables, such as the quadrupole moment, are difficult to obtain 
using effective operators consistent with the interaction \cite{Stetcu:2004wh}.
The description of low-momentum 
observables requires large model spaces, 
which makes the application of the method to heavy nuclei extremely 
challenging.

These problems can be mitigated if one formulates the
problem as an EFT in a discrete basis.
The basic idea underlying EFTs is that the restriction of
a theory to
a model space generates all interactions allowed by the theory's symmetries
\cite{kaplan}.
Since particle momenta are limited within the restricted space,
one can treat short-distance interactions in a derivative expansion,
similar to the multipole expansion in classical electrodynamics.
The coefficients of this expansion carry information
about the details of the short-range dynamics.
In addition, these parameters depend on the size of the model space
in such a way as to make the results for low-energy observables independent of
the size of the space. 
Even in the absence of exact solutions from QCD, EFTs 
provide a modern understanding of the nuclear forces at low energies
\cite{eftreview}.
An EFT with pion fields can be constructed and generalizes
chiral perturbation theory to nuclear systems,
although its renormalization is still being explored \cite{nogga}.
At sufficiently low energies, a 
simpler EFT without explicit pions exists,
and it has been shown to have a well-defined organizational principle 
(``power counting'') \cite{pionless},
and give 
good results for 
$^2$H \cite{rupak},
$^3$H \cite{triton_eft}, and even the ground state of 
$^4$He \cite{Platter:2004zs}. 
In leading order (LO), the pionless EFT has three 
parameters: two NN contact interactions that contribute
to the NN $^3S_1$ and $^1S_0$ channels \cite{pionless,rupak}, 
and one 3N contact interaction
that appears in the 3N $S_{1/2}$ channel \cite{triton_eft}.
With trivial modifications, this EFT has also proved 
useful for atomic and molecular systems with
large scattering lengths \cite{hammerrev}.
While continuum momentum-space calculations are  considerably  
simplified in this EFT,
they are still quite involved beyond the four-body system. 
First attempts have been made to derive bulk properties of 
matter using a spatial lattice \cite{lattice},
but the limit of applicability of the pionless theory with increasing density
is at present unknown.

In this paper, we combine the virtues of the two approaches: we explore
an implicit removal of the so-called excluded (or $Q$) space in the NCSM 
by using the EFT principles (including power counting), 
thereby providing solutions of the EFT for heavier systems.  
The earlier successes of the standard shell model 
(an inert core plus a few valence nucleons) suggest that it might be
interpretable as an EFT.
For example, one hopes that EFT power counting might eventually
explain why a good 
description of nuclear
properties can be achieved using essentially one- and two-body
terms in the Hamiltonian plus some $A$-dependence \cite{wildenthal}.
While some attempts to infuse EFT ideas into the shell model have been
investigated in the past \cite{haxtonSMasEF}, we follow a different
approach.

We start with the EFT Hamiltonian in the NCSM basis.
The HO frequency, which we denote by $\omega$,
sets the spacing between HO levels and provides an infrared energy cutoff
$\hbar\omega$
or, equivalently, a momentum cutoff 
$\lambda=\sqrt{m_N\hbar\omega}$, where $m_N$ is the nucleon mass.
For the two-body problem one defines a truncation of the model
space that includes 
all the HO states with the quantum number $N=2n+l\le N_{max}$,
where $n$ ($l$) denotes the radial (angular-momentum) quantum number. 
Because we investigate here positive-parity states only, we limit
$N_{max}$ to even values.
The usual definition of an ultraviolet cutoff $\Lambda$ 
in the continuum 
can be extended to discrete HO states. Thus, in the HO basis, 
we define $\Lambda=\sqrt{m_N(N_{max}+3/2)\hbar\omega}$ as the 
momentum associated with the energy of the highest HO level. 
A model space is defined by the two cutoffs.
Within each model space 
we fix the parameters of the EFT to reproduce
a set of observables and  then calculate other observables 
in the same spaces.
Because we construct the effective interaction directly in the model space, 
no additional unitary transformation is required, as it is 
in the conventional 
NCSM approach \cite{Navratil:2000ww}.
Contrary to the latter, here  the effective interaction does not
have to converge to an underlying ``bare'' interaction.
What is important 
is the behavior of observables with 
the variation of the ultraviolet cutoff $\Lambda$. 
In each order in the power counting one expects independence of 
the physical observables with respect to 
$\Lambda$, within the theoretical errors induced by the absence of the 
higher-order operators that come at the next orders. 
There might also be a 
dependence on the infrared  cutoff $\lambda$,
which
is an artifact of using a HO basis and should be removed. 
The infrared cutoff in this case is analogous to the 
inverse of the lattice size in a lattice discretization \cite{Seki:2005ns}. 
Obviously, as $\Lambda=\lambda\sqrt{N_{max}+3/2}$, the running of the 
observables with $\Lambda$ 
cannot 
be obtained by  increasing $\hbar\omega$ arbitrarily 
in a fixed-$N_{max}$ model space, 
as this procedure increases the infrared cutoff as well, 
introducing additional errors. 
Instead, we verify explicitly that cutoff dependences decrease 
with increasing $\Lambda$
and decreasing $\lambda$, and 
we remove the influence of the 
infrared cutoff by 
extrapolating to the continuum limit,
where $\hbar\omega\to 0$ with $N_{max}\to\infty$ 
so that $\Lambda$ is fixed. 
Traditional shell-model calculations 
use larger values for the HO frequencies, 
of the order of $41/A^{1/3}$ MeV, but
in our approach, we are interested in a small infrared cutoff limit, 
which removes the HO frequency dependence.
At the end, we also extrapolate to the $\Lambda\to \infty$ limit.

Our method can be applied to either pionful or pionless EFTs.
For simplicity, in this first application 
we limit ourselves to the pionless EFT in LO.
The interaction in each model space has the same structure, 
\textit{i.e.}, matrix elements of the 
contact two- and three-body interactions.
The strengths of the three coupling constants
have to be adjusted to reproduce three experimental observables, 
which here we choose to be the energy levels of the lightest nuclei:
in the NN $^3S_1$ channel we use the $^2$H 
binding energy to determine one coupling constant, 
while the other NN coupling constant (in the $^1S_0$ channel) 
and the strength of the contact 3N term 
(in the $S_{1/2}$
channel) are determined so that one simultaneously describes correctly 
the $^3$H and 
$^4$He binding energies. 

The three- and four-body problems are solved using a relative-coordinate 
code developed following Ref. \cite{Navratil:1999pw}. For four particles,
the largest model space we can handle has 
$N_{max}=16$.
The calculations are much more involved for $A\ge5$, because the 
relative-coordinate approach becomes inefficient
due to the extremely difficult anti-symmetrization. 
To overcome this issue, 
we use a Slater-determinant (SD) basis constructed from HO
single-particle states. 
The contact NN and 3N interactions in the $S$ channels are exactly
transformed to a proton-neutron SD basis 
\cite{PhysRevC.3.1137},
and the many-body diagonalization is performed with the shell-mode code 
{\sc redstick} \cite{redstick}.
For $^6$Li, investigated in this paper, the largest model space 
that we can handle when including 
3N forces has $N_{max}=8$, or, equivalently, 
$6\hbar\omega$ excitations allowed above the lowest configuration.
(Note that a complete $N\hbar\omega$ basis allows an exact separation 
of the spurious center-of-mass excitations, although the SD basis is 
not translationally invariant.) 

\begin{figure}
\includegraphics*[scale=0.8]{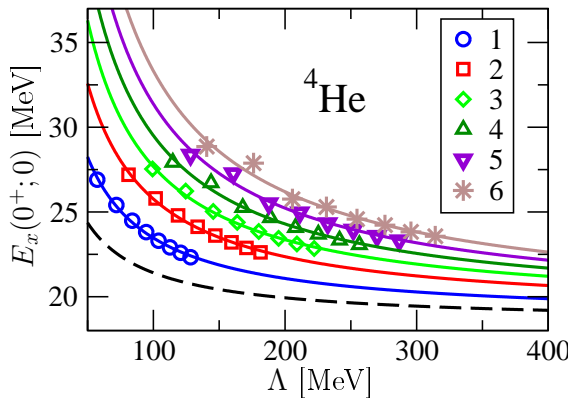}
\caption{(Color online) 
Dependence of the first $(0^+;0)$ excitation energy
in $^4$He on the ultraviolet momentum cutoff $\Lambda$. 
For each frequency $\hbar\omega$, given in the legend in MeV, we interpolate 
the direct results (discrete symbols) with a $1/\Lambda$ dependence 
(continuous curves).  The dashed curve marks the limit $\hbar\omega\to 0$
(see text for details).
\label{He4_fx_Nmax}}
\end{figure}

With the deuteron, triton and alpha-particle ground-state energies 
used as input to determine the effective interaction, 
the only energy levels that we can predict for $0s$ nuclei 
are the excited states of $^4$He,
which have not previously been calculated in the pionless EFT. 
In Fig. \ref{He4_fx_Nmax}, we show the results for the excited-state energy 
of the first spin-parity $J^\pi=0^+$, isospin $T=0$, shorthanded $(0^+;0)$, 
state in $^4$He, as a function of the cutoff momentum $\Lambda$.  

As we are interested in the limit of large $\Lambda$ and small $\lambda$,
and very large $N_{max}$ calculations are prohibitive even for few particles, 
we extrapolate the results obtained at smaller $N_{max}$ by fitting our results
to $E_0(\hbar\omega)+A(\hbar\omega)/\Lambda$, as shown by the continuous lines 
in Fig. \ref{He4_fx_Nmax}. 
The choice of the $1/\Lambda$ dependence is motivated by the same type of 
LO running of the bound-state energy in the two-body sector 
in the continuum \cite{pionless}.
Although we cannot exclude the appearance of a softer dependence on $\Lambda$
due to discrete-space and many-body effects,
we obtain a good fit with this simple formula.
(As a check, we found no significant change when we
added a term $\log(a_0\Lambda)/\Lambda$ to the fit.
More extensive calculations might be able to further
pin down the running for the many-body system.)

\begin{figure}
\includegraphics*[scale=0.77]{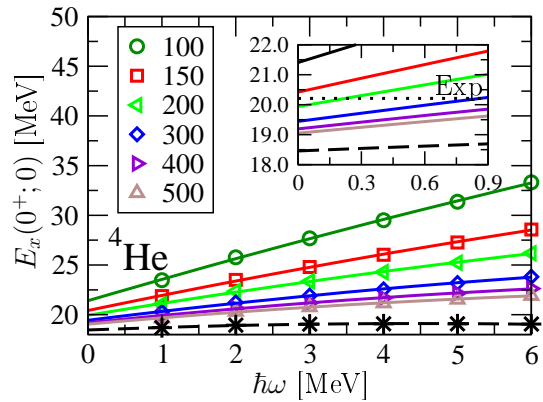}
\caption{(Color online) 
Dependence of the first $(0^+;0)$ excitation energy in $^4$He 
on the infrared energy cutoff $\hbar\omega$.
For each ultraviolet cutoff $\Lambda$, given in the legend in MeV, 
we interpolate as described in the text. The results marked with 
star symbols are obtained in the limit $\Lambda\to\infty$.
In the insert we show the variation around origin,
compared to the experimental value.
\label{He4_fx_hw}}
\end{figure}

We use the extrapolation above to study the $\lambda$ dependence of
the first $(0^+;0)$ energy.
In Fig. \ref{He4_fx_hw}, we present the results 
for selected ultraviolet-cutoff 
values, and also show the $\Lambda\to \infty$ limit, $E_0(\hbar\omega)$.
Figure \ref{He4_fx_hw} shows a simple dependence upon the HO frequency. 
Thus, we interpolate the results for fixed $\Lambda$ with second-degree 
polynomials.
The coefficients of the squared dependence come out very small, 
so that the dependence is almost linear. 
The constant terms in the quadratic interpolation are then 
fitted ($\chi^2\approx 10^{-4}$) with $E_0(0)+A(0)/\Lambda$.
The resulting ultraviolet-cutoff dependence is shown
as the dashed line in  Fig. \ref{He4_fx_Nmax},
and represents the continuum limit $\hbar\omega\to 0$.
We find $E_0(0)=18.5$ MeV, in remarkable (for a LO calculation)
agreement
with the experimental value  $20.21$ MeV \cite{A4exp}.
We can perhaps understand this agreement
if we consider that this state is very close to the 
four-nucleon continuum threshold, 
in an energy regime well within the validity of the pionless EFT.

\begin{figure}
\includegraphics*[scale=0.7]{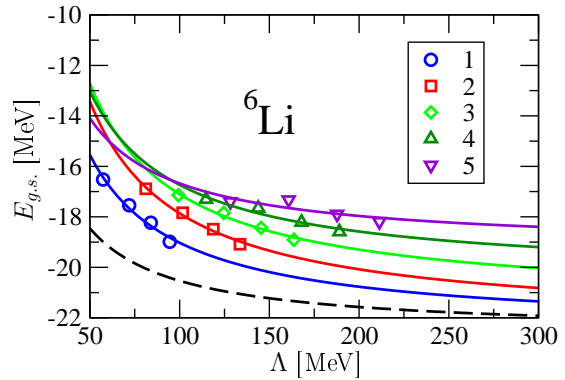}
\caption{(Color online) Same as in Fig. \ref{He4_fx_Nmax}, 
but for the ground-state energy of $^6$Li. 
\label{Li6gs}}
\end{figure}

Following the same procedure,
the ground-state energy of 
$^6$Li is evaluated, also for the first time in the pionless EFT. In this case,
we obtain simultaneously several low-lying positive-parity states, 
irrespective of the
spin or isospin values.
Some $T=0$ states appear degenerate in our calculation, 
which is not necessarily surprising, given that the spacing between levels 
is less than the expected error in the pionless EFT. 
However, we always obtain a single $(1^+;0)$ ground state, 
as found experimentally,
its energy as function of $\Lambda$ being given in Fig. \ref{Li6gs}.
We remove the infrared-cutoff dependence 
using the same procedure as for  $^4$He.
Although the fitting is subject to larger errors because of the 
limitation in $N_{max}$,
and the quadratic dependence upon $\hbar\omega$ is more pronounced 
in the case of $^6$Li, we obtain again 
a $\Lambda^{-1}$ running in the continuum limit,
as shown by the dashed curve in Fig. \ref{Li6gs}.
For large values of the ultraviolet cutoff $\Lambda$, we estimate
the binding energy as about 22.6 MeV,
to be compared to the experimental result,
31.99 MeV \cite{expA5-10}. 
While not as precise as the excited-state energy in $^4$He, our result
for the $^6$Li binding energy is consistent with the expected LO errors 
in the pionless EFT.
 
We expect agreement to improve in higher orders, as in continuum calculations
\cite{eftreview}.
In LO, we were forced to use very large 
model spaces, restricting applicability to light systems. 
However, 
higher orders should produce reduced cutoff errors
(that is, smaller coefficients of the $1/\Lambda$ term),
faster convergence, and thus extended applicability to heavier systems.
Since parameters multiply in subleading orders, 
a method to fit them to scattering
data is desirable.
However, 
because the basis states in the NCSM 
are constructed using HO wave functions, 
their bound-state asymptotics do not allow a direct determination of 
scattering observables. 
We are currently in the process of developing an approach 
to 
calculate scattering observables using a 
truncated HO basis \cite{HOphsh}.

Note that our results do not include Coulomb contributions explicitly.
One-photon exchange between nucleons is non-perturbative only
for momenta below about $\alpha m_N$, where $\alpha$ is the
fine-structure constant. Since nucleons in the bound states
we consider have much larger typical momenta, electromagnetic
interactions appear only in subleading orders.
We thus used the observed $^4$He binding energy as a fitting
parameter; the difference between that and a Coulomb-corrected value
is a higher-order effect.
Of course, since they grow with the 
square of the 
number of protons, Coulomb effects are larger in heavier systems.
In particular, we expect our $^6$Li results to improve somewhat
when we include Coulomb explicitly. 
This is because in our current calculation 
the Coulomb repulsion in $^4$He  is being divided through
the number of nucleon pairs, and therefore implicit Coulomb effects are
growing roughly as the square of the number of nucleons,
rather than the number of protons as when Coulomb is included
explicitly.
A simple estimate suggests that this results in an
underbinding of about 2 MeV in the ground-state energy of $^6$Li.
This 10\% effect is, as expected, well in line with the size of
subleading orders, but in itself would move the  $^6$Li
energy to within 25\% of the observed value.

In summary, we have investigated a new \textit{ab initio} method for 
solving many-body problems in truncated model spaces. 
While our main application is to the description of 
properties 
of light nuclei, the principles are general, allowing similar treatment 
of other interacting quantum many-body systems. 
In our approach, the effective interaction is directly determined in 
the model space, where an exact diagonalization in a complete many-body basis 
is then performed.  
Power counting is seen as the justification
for a cluster approximation.
Our LO prediction for the energy of the first $(0^+;0)$ excited state 
in $^4$He 
is within $10\%$ of the experimental value, while the $^6$Li ground-state 
energy is predicted at about 70\% of the experimental value. 
These results are encouraging, and suggest that
the method can be improved systematically by going to higher orders.
The result in the six-nucleon system is considerably worse than
for lighter systems, but is not in disagreement with the expected
error in the pionless EFT at LO.
Thus, more investigation is needed before any definitive
conclusion can be drawn with regards to a possible breakdown of the 
pionless theory in heavier systems. 
In any case, applying the same procedure to the pionful EFT
should increase the range of applicability of the method.
Although we have not yet discussed other observables, the approach presented 
in this paper also offers the possibility of a consistent treatment of 
external operators on the same footing with the effective interaction, 
following the same general EFT principles.\\

\noindent
{\it Acknowledgments.}
We thank W.E. Ormand for providing the three-body version of the shell-model 
code {\sc redstick}, and C.W. Johnson and J.P. Vary for useful discussions.
I.S. and B.R.B. acknowledge partial support by NSF grant numbers PHY0244389
and PHY0555396. U.v.K. acknowledges partial support from DOE grant number 
DE-FG02-04ER41338 and from the Sloan Foundation. 


\end{document}